\documentclass[aps,prl,twocolumn,showpacs,superscriptaddress,groupedaddress]{revtex4}  
\usepackage{graphicx}  
\usepackage{dcolumn}   
\usepackage{bm}        
\usepackage{amssymb}   
\usepackage{ulem}
\usepackage{color}

\begin{document}

\title{Dynamic nuclear polarization from current-induced electron spin polarization}

\author{C. J. Trowbridge}\affiliation{Department of Physics, University of Michigan, Ann Arbor, MI 48109}
\author{B. M. Norman}\affiliation{Department of Physics, University of Michigan, Ann Arbor, MI 48109}

\author{Y. K. Kato}\affiliation{Institute of Engineering Innovation, University of Tokyo,Tokyo 113-8656, Japan}

\author{D. D. Awschalom}\affiliation{Institute for Molecular Engineering, University of Chicago, Chicago, IL 60637}

\author{V. Sih}\affiliation{Department of Physics, University of Michigan, Ann Arbor, MI 48109}
\date{\today}

\begin{abstract}
Current-induced electron spin polarization is shown to produce nuclear hyperpolarization through dynamic nuclear polarization.  Saturated fields of several millitesla are generated upon the application of electric field over a timescale of a hundred seconds in InGaAs epilayers and measured using optical Larmor magnetometry.  The dependence on temperature, external magnetic field, and applied voltage is investigated.  We find an asymmetry in which the saturation nuclear field depends on the relative alignment of the electrically generated spin polarization and the external magnetic field, which we attribute to an interplay between various electron spin dynamical processes.    
\end{abstract}

\maketitle

The nuclear spin system in semiconductors has attracted interest for potential applications in classical and quantum spin-based computation schemes \cite{epstein_2005, reimer_2010,xu_2009}.  Its isolation from the surrounding environment yields exceptionally long coherence times, which can be as much as nine orders of magnitude longer than electron spin coherence times \cite{kikkawa_2000}, and suggests use as an intermediate timescale data storage mechanism \cite{mccamey_2010}.  For magnetic resonance imaging, large magnetic fields are required to produce a sufficient number of spins for a detectable signal \cite{mahron_1995}.  Both imaging and information processing applications stand to benefit from methods for controlling and exceeding the equilibrium nuclear spin polarization. 

Dynamic nuclear polarization (DNP) has been shown to generate nuclear polarizations which exceed the equilibrium value.  Through DNP, which occurs when electron spins that have been driven out of thermodynamic equilibrium attempt to thermalize through hyperfine coupling to the nuclear spin system, the nuclear spin system can be manipulated indirectly through control of the electron spin system.  This was first achieved by using microwave fields to saturate electron spin resonance (Overhauser and solid effect) \cite{overhauser_1953a,overhauser_1953b}.  It has since been demonstrated by generating a non-equilibrium electron spin polarization by optical pumping \cite{lampel_1968,paget_1977,kikkawa_2000}, ferromagnetic imprinting \cite{kawakami_2001}, electrical spin injection from a ferromagnet \cite{epstein_2003,strand_2003}, and in a spin-polarized Landau level \cite{hashimoto_2002,li_2012,lo_2013}.

Methods that use electric fields have the advantage that they can be applied more locally.  In 1959, Feher proposed a hot electron effect, in which a dc electric field is used to increase the electron temperature relative to the nuclei \cite{feher_1959}.  This effect is analogous to the radiofrequency field used in the Overhauser effect and was demonstrated in InSb by Clark and Feher in 1963 \cite{clark_1963}.  More recently, current-induced dynamic nuclear polarization experiments which rely on the hot electron effect were conducted in GaAs \cite{hoch_2005} and InP \cite{kaur_2010}.  Our results demonstrate another mechanism by which current can enhance nuclear polarization, which is through the electron spin polarization generated by current-induced spin polarization (CISP) \cite{kato_2004,norman_2014}.  Here, the direction of the current with respect to the crystal axes determines the magnitude and direction of the electrically-generated electron spin polarization and the resulting nuclear spin polarization.  The magnetic field due to nuclear polarization is then measured via optical Larmor magnetometry \cite{kikkawa_2000,salis_2001}.

In our experiment, DNP occurs through the contact hyperfine interaction between the lattice nuclei and itinerant conduction band electrons and/or those trapped by shallow donor sites and impurities.  The coupling between the nuclear spin system and the fluctuating hyperfine field resulting from the electron spin magnetic moments leads to nuclear spin polarization decay with lifetime $T_{1e}$ \cite{abragam}. If the electrons are maintained out of equilibrium by some means of pumping, this mechanism results in DNP. At equilibrium and neglecting thermal electron alignment, the average nuclear spin polarization can be expressed as \cite{optical_orientation}: 
\begin{equation}
\vec{I}_{av} = \frac{4}{3}I(I+1)\frac{(\vec{B}\cdot\vec{S})\vec{B}}{B^2}
\end{equation}
where $I$ is the total spin of the nuclei and $\vec{S}$ is the average electron spin. The nuclear polarization in turn gives rise to a magnetic field,  given by $\vec{B}_N = \sum_\alpha \vec{I}_{av} b_{N,\alpha}f_\alpha/I_\alpha$, where the sum is over the nuclear species, $b_{N,\alpha}$ is the field from complete saturation of species $\alpha$ and $f_\alpha$ is a species-dependent �leakage factor� given by $T_1/(T_1+T_{1e})$ where $T_1$ is the nuclear relaxation time due to other channels.  Previous measurements have shown that the degree of electron spin polarization attained by CISP has an upper bound of order $10^{-3}$ in our samples with our experimental parameters \cite{norman_2014,trowbridge_2011}. Accordingly, we expect a nuclear field on the order of 1-10 mT in our system assuming $T_1$ is long compared to $T_{1e}$.

All samples used in this study consist of a 500 nm thick layer of Si-doped $n=3\times 10^{16}\  $ cm$^{-3}$ In$_{.04}$Ga$_{.96}$As grown by molecular beam epitaxy atop semi-insulating $[001]$ GaAs substrate, and capped with 100 nm of GaAs.  Ohmic contacts are deposited to drive in-plane current.  Samples A and B have four contacts around a square mesa-etched region designed so that a current can be driven in any in-plane direction \cite{meier_2007,norman_2014}.  Further details of the sample design can be found in Ref. \cite{norman_2014}.  Numerical calculations find a region of electric field uniformity with a radius of 35 $\mu$m in which the amplitude deviates by less than 5\% and its direction by less than 5 degrees. The pump and probe beam diameters were measured to be 30-35 $\mu$m.  Errors in placing the beam at the center of the sample could introduce errors in the electric field amplitude and direction.  For sample C, 400 $\mu$m long by 100 $\mu$m wide channels were etched along the $[110]$ and $[1\bar{1}0]$ crystal directions.  This sample design allows for a higher electric field for a given power dissipation and errors in electric field direction and magnitude due to beam placement are eliminated.  However, measurements for different crystal directions are performed on different channels, and previous measurements have shown that the spin-orbit field and CISP magnitudes vary strongly with position, perhaps as a result of inhomogeneous uniaxial strain \cite{norman_2010, norman_2014}.

\begin{figure}
\includegraphics[scale=1]{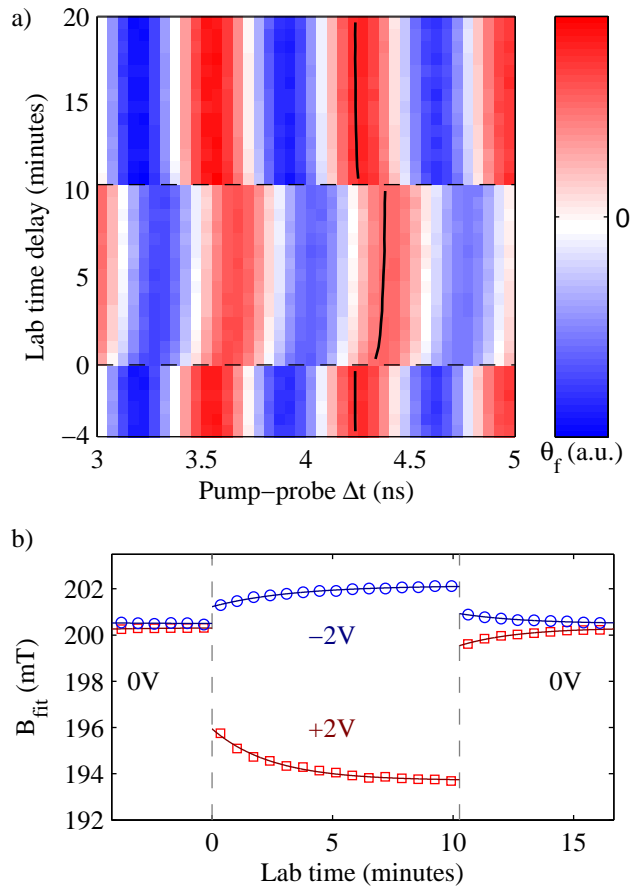}
\caption{\label{fig:epsart} a) Series of Faraday rotation time delay scans showing a transition from $V_{DC} = 0$ V  to  2 V at lab time  0 and back to 0 V after 10 minutes .  Data were taken on sample B with current flowing along $[1\bar{1}0]$ at 10 K with 200 mT external field applied.  Solid black line indicates position of local maximum from fits to Faraday rotation signal.  b) Total magnetic field as measured from fits to delay scans shown in a) (red squares) along with another similar transition to $V_{DC} = -2$ V (blue circles).  Lines show exponential fits to magnetic field data.  Fits allow extraction of saturation nuclear field $B_N$ and saturation time $T_{1e}$.}
\end{figure}

The sample is placed in a helium flow cryostat which is held between the poles of an electromagnet.  A magnetic field is applied in the sample plane, both to suppress nuclear spin relaxation by magnetic dipole-dipole interactions \cite{abragam} and to perform optical Larmor magnetometry. In Larmor magnetometry, test electron spins are optically injected using a circularly polarized pump pulse and their precession about the total magnetic field is monitored by time resolved Faraday rotation \cite{kikkawa_2000,kawakami_2001}. The circularly polarized pump beam is modulated between left and right circular polarization at a frequency of 50 kHz by a photo-elastic modulator for lock-in detection.  The external magnetic field causes these test electron spins to precess at a high enough frequency that many rotations can be measured over the time delays accessible to the mechanical delay line. This allows for measurement of the total magnetic field about which the electrons precess to be measured to a precision of approximately 100 $\mu$T in the 40 seconds it takes to complete a scan of the pump-probe delay time. 

\begin{figure}
\includegraphics[scale=1]{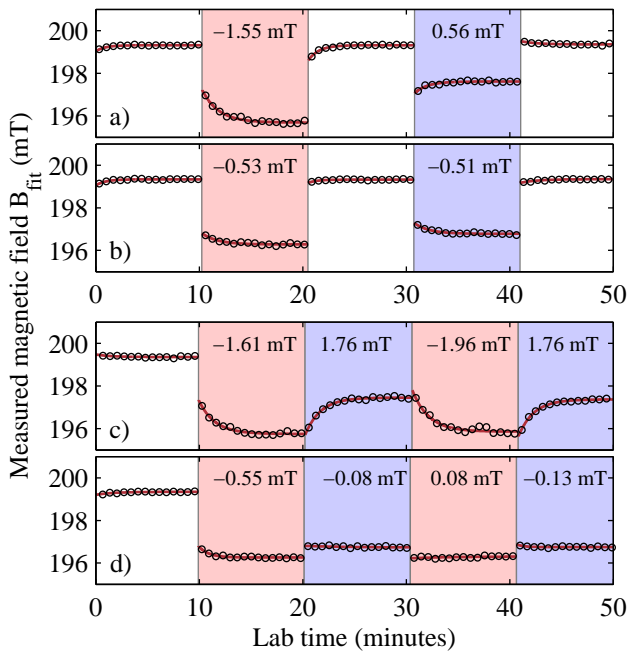}
\caption{\label{fig:epsart} Total magnetic field measured via Larmor magnetometry following voltage transitions with CISP parallel (parts a) and c)) or perpendicular (parts b) and d)) to the external magnetic field.  Red and blue shading indicate $V_{DC} = 2$ V and -2 V, respectively. Inset text indicates total change in nuclear field $\Delta B_{N}$ in the labelled transition.  Plots a) and b) show transitions of the form $V_{DC} = 0\rightarrow \pm 2$ V; the observed asymmetry with CISP parallel to $B_{ext}$ in a) results from current direction-independent DNP mechanisms, which are seen in b) when CISP is perpendicular to $B_{ext}$. By considering transitions of the form $V_{DC} = \pm 2 \rightarrow \mp 2$ V (plots c) and d)), contributions to $\Delta B_{N}$ from current direction-independent mechanisms are suppressed, isolating DNP due to CISP and highlighting the strong directional dependence of $\Delta B_N$ due to CISP on current direction.}
\end{figure}

\begin{figure}
\includegraphics[scale=0.5]{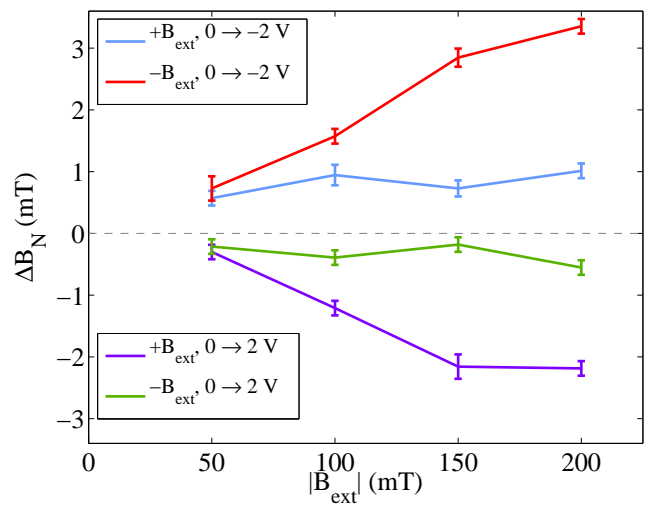}
\caption{\label{fig:epsart} Saturation nuclear field versus applied magnetic field for four different types of voltage transitions (described in figure legend), showing asymmetry of unipolar transition saturation amplitudes.  Measurements were taken on sample B with current along $[1\bar{1}0]$ at 10 K.  Red and purple data sets show strong dependence on external field and correspond to a geometry in which the nuclear alignment and external magnetic field are antiparallel.  Blue and green data sets correspond to nuclear alignment parallel to external field.}
\end{figure}

Figure 1 shows a measurement performed on Sample B with current along $[1\bar{1}0]$ and at a temperature of 10 K.  Fig. 1 a) contains a plot of the Faraday rotation signal observed due to the test electron spin packet as a function of pump-probe time delay (horizontal axis) and lab time (vertical axis).  At lab time 0, a voltage is applied across the sample.  A rapid shift in the precession frequency, corresponding to a change in field of a few millitesla, occurs due to the spin-orbit field \cite{meier_2007}.  A slow shift in the precession rate follows, which we attribute to nuclear polarization.  After 10 minutes, the voltage is switched off and the nuclear spin polarization decays.  Each time delay scan is fit to extract the electron Larmor precession frequency, given by $\Omega_L = g\mu_BB/\hbar$, and the total field about which the electrons precessed is calculated. These results are plotted in Fig. 1 b), along with a fit to the equation $B(t_L) = \Delta B_N(1-\exp[-t_L/T_{1e}]) + B_{\text{0}}$ where $t_L$ is lab time and $B_{\text{0}}$ is the sum of the external and spin-orbit fields.  The saturation change in nuclear field $\Delta B_N$ and the polarization time $T_{1e}$ are extracted from the fit. The transition from V$_{DC} = 0 \rightarrow 2$ V shows $\Delta B_{N} = -2.2$ mT and $T_{1e} = 148$ s, while the transition from V$_{DC} = 0 \rightarrow -2$ V shows $\Delta B_{N} = 1.0$ mT  and $T_{1e} = 198$ s. 

There is a readily apparent asymmetry shown in Fig. 1; the transition to +2 V shows a larger shift in nuclear field than the transition to -2 V.  The origin of this asymmetry is investigated in Fig. 2, in which transitions in two different geometries are shown.  Measurements are taken on sample B at 10 K with a 200 mT external magnetic field and current along $[1\bar{1}0]$.  Here, red and blue shading indicates $V_{DC} = 2$ or -2 V, respectively, and the inset text shows the measured values of $\Delta B_N$ for each labelled transition. Plot a) shows a set of transitions with CISP oriented parallel to the external magnetic field $B_{ext}$, while in plot b) CISP is perpendicular to $B_{ext}$.   With CISP perpendicular to $B_{ext}$, the $(\vec{B}\cdot\vec{S})$ term in Eq. 1 suggests that there should be no observable DNP, however a non-zero $\Delta B_N$ is measured.  Here, the direction of the current does not significantly alter the observed $\Delta B_N$.  

The observed $\Delta B_N$ with CISP perpendicular to $B_{ext}$ can be understood as resulting from the hot electron effect \cite{feher_1959} and/or the presence of the pump and probe beams \cite{kikkawa_2000}. The hot electron effect results in a heating of the electron spin system which varies with the magnitude of current in the sample but not its direction. Additionally, the pump and probe beams, which are tuned just below the absorption edge, result in  photo-excited carriers which are nominally unpolarized in the axis of quantization defined by the external magnetic field in the Voigt geometry. These optically injected spins result in heating of the electron spin system where they are present.  When a voltage is applied, photo-excited carriers will be driven out of the region of interrogation, giving rise to a voltage-dependent change in nuclear spin polarization which would depend on the voltage magnitude and absorbed pump and probe power.  Further measurements are required to quantify the contribution of each mechanism. The asymmetry between transitions to +2 V versus -2 V seen in Fig. 2(a) can then be explained as the result of an interplay between DNP due to CISP and DNP due to isotropic mechanisms outlined above.  By subtracting the values of $\Delta B_N$ observed in transitions with CISP perpendicular to $B_{ext}$ from those with CISP parallel to $B_{ext}$, the contribution to $\Delta B_N$ from CISP is isolated, and the asymmetry disappears.  

By considering transitions of the type $V_{DC} = \pm$V $\rightarrow \mp$V after saturation at V, contributions to changes in nuclear field caused by mechanisms which do not depend on the direction of current are suppressed, allowing for current direction-dependent alignment mechanisms to be studied in isolation. Figures 2 c) and d) show measurements where these transitions are performed with CISP parallel and perpendicular to $B_{ext}$, respectively, at lab times of about 20, 30, and 40 minutes. These measurements highlight the strong dependence of $\Delta B_N$ on the orientation of the current in the sample; $\Delta B_N$ with CISP parallel to $B_{ext}$ is an order of magnitude larger than $\Delta B_N$ with CISP perpendicular to $B_{ext}$. 

The behavior of $\Delta B_N$ with external magnetic field and the sign of the applied voltage is shown for Sample B at 10 K and with current along the $[1\bar{1}0]$ direction in Fig. 3.  Reported error bars represent the standard error of a set of 6 measurements at each point.  These data reflect the asymmetry discussed above. Transitions with B$_{ext}$ antiparallel to the change in nuclear field (red and purple) lead to a larger measured $\Delta B_N$ at our experimental parameters than transitions in which the nuclear field and external field are parallel (blue and green). This asymmetry remains consistent with a reversal of the direction of the external magnetic field, so that in each case the transition whose nuclear field is changing so as to oppose the external magnetic field results in a larger $\Delta B_N$.

\begin{figure}
\includegraphics[scale=1]{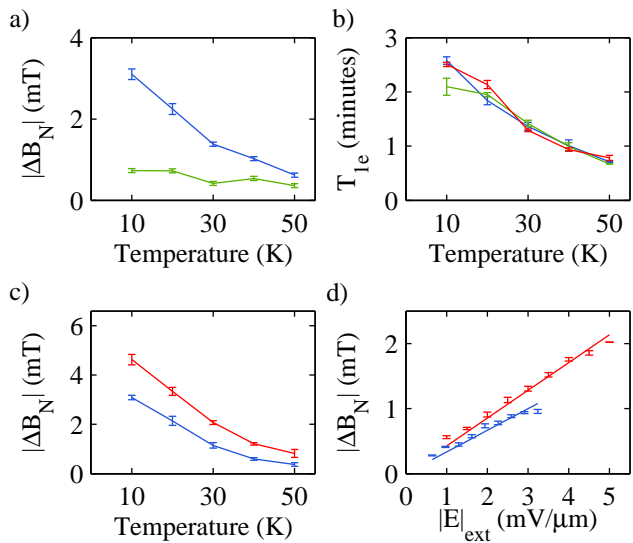}
\caption{\label{fig:epsart}  a) Magnitude of $\Delta B_N$ as measured on sample B with $B_{ext} = 200$ mT for transitions of $V_{DC} = 0 \rightarrow 2$ V (blue) and $V_{DC} = 2 \rightarrow 0$ V.  b) $T_{1e}$ measured as a function of temperature on sample B with $B_{ext} = 200$ mT for transitions $V_{DC} = 0 \rightarrow 2$ V (blue), $V_{DC} = 2 \rightarrow 0$ V (green) and $V_{DC} = \pm2 \rightarrow \mp2$ V (red).  Though all relaxation times are similar, the scaling with temperature of $\Delta B_N$ in transitions of $V_{DC} = 2 \rightarrow 0$ V show unexpected scaling with temperature. c) $B_N$ vs. temperature using sample C with current along $[110]$ for 2 $ \leftrightarrow$ -2 V (red) and 1 $ \leftrightarrow$ -1 V (blue) transitions. Sublinear scaling with voltage below 30 K indicates onset of sample heating. d) $|\Delta B_N|$ vs. applied electric field $|E_{ext}|$ for samples B (blue) and C (red), along with linear fits to the data. Data were taken from $\pm 2$ $\leftrightarrow\mp 2$ V transitions at 30 K with a 200 mT external field.  }
\end{figure}

Measurements of $\Delta B_N$ and $T_{1e}$ as a function of sample temperature are shown in Figures 4 a) and b), respectively. In Fig. 4 a), the blue curve shows $\Delta B_N$ versus temperature for transitions of the form 0 V $\rightarrow$ 2 V while the green curve shows the opposite transition 2 V $\rightarrow$ 0 V.  These two transitions show strikingly different behaviors of $\Delta B_N$ with temperature.  However, as shown in Fig. 4 b), the timescale over which these transitions occur is similar.  This suggests that there is another mechanism, which must take place at a time scale faster than is accessible by our experimental method, that is responsible.  This may take the form of a rapid dynamic process which occurs when the applied voltage is changed.  

Figure 4 c) shows the behavior of $|\Delta B_{N}|$ with temperature for $\pm$2 V $\leftrightarrow \mp$2 V (red) and $\pm$1 V $\leftrightarrow\mp$1 V (blue) transitions.  Measurements were attempted at 60 K; while the precession of the test electron spin packet was visible, no $\Delta B_N$ was seen.  This is consistent with the behavior of the nuclear field found in previous measurements \cite{kikkawa_2000}.  At temperatures below 30 K, the signal does not  double with the magnitude of the voltage.  This can be attributed to heating in the sample, which is of most concern below 30 K, due in part to a decrease in the thermal conductivity of the GaAs substrate \cite{carlson_1965}.  Figure 4 d) shows the saturated nuclear field strength at 30 K with $B_{ext} = 200$ mT for sample B with current along $[1\bar{1}0]$ (blue) and sample C with current along $[110]$ (red) as a function of the applied electric field.  The design of Sample C allows for higher applied electric fields at a given thermal power dissipation.  That the saturated nuclear field scales linearly with the applied electric field agrees with previous measurements of the degree of electron spin polarization due to CISP in these samples \cite{norman_2014,kato_2004,trowbridge_2011}.  This result was found to be consistent on all samples and both orientations used in this study.  In addition, the slopes were consistent with independent measurements of CISP strength, as expected \cite{norman_2014}.

We have performed measurements of DNP due to CISP using Larmor magnetometry in n-InGaAs.  Nuclei in the material are polarized in a direction which is determined by the electron spin polarization due to CISP.  Changes in magnetic field due to nuclear polarization are measured as temperature, applied voltage, orientation, and applied magnetic field are changed, and are found to be as large as a few millitesla in the range of experimental parameters used, which corresponds to fields more than an order of magnitude larger than thermal polarization. We find an asymmetry in the scaling of the saturation nuclear field for differing current and magnetic field directions, which can be attributed to competing electron spin dynamical processes.  Future work should focus on quantifying the role of identified mechanisms as sources of asymmetry found here, as well as the rapid depolarization upon removing the DC voltage.  

This material is based upon work at Michigan supported by the National Science Foundation under Grant No. ECCS-0844908 and the Materials Research Science and Engineering Center program DMR-1120923, the Office of Naval Research, the Air Force Office of Scientific Research, and the Defense Threat Reduction Agency, Basic Research Award \#HDTRA1-13-1-0013. Sample fabrication was performed in the Lurie Nanofabrication Facility, part of the NSF funded NNIN network. The work at Chicago is supported by the National Science Foundation and the Office of Naval Research.

\end{document}